\documentclass[preprint,12pt]{elsarticle}
\usepackage[utf8]{inputenc}
\usepackage{amsfonts,amsmath,amssymb}
\usepackage{graphicx}
\usepackage{hyperref}
\usepackage{color}
\journal{Physics of the Dark Universe}

\def\Lcdm{\Lambda {\rm CDM}}
\def\Mpl{M_{\rm pl}}
\def\H0{\mathcal{H}_{0}}

\begin{document}

\begin{frontmatter}

\title{The parameter space of Cubic Galileon models for cosmic acceleration}
\author[n,nn,nnn]{Emilio Bellini}
\author[nnnn,nnnnn]{Raul Jimenez}
\address[n]{Institut f\"{u}r Theoretische Physik, Universit\"{a}t Heidelberg, Philosophenweg 16, 69120 Heidelberg, Germany}
\address[nn]{Dipartimento di Fisica e Astronomia ``G. Galilei'', Universit\`a degli Studi di Padova, via Marzolo 8, I-35131, Padova, Italy}
\address[nnn]{INFN, Sezione di Padova, via Marzolo 8, I-35131, Padova, Italy}
\address[nnnn]{ICREA \& ICC, University of Barcelona (UB-IEEC), Marti i franques 1, 08034, Barcelona, Spain}
\address[nnnnn]{Theory Group, Physics Department, CERN, CH-1211, Geneva 23, Switzerland}

\begin{abstract}
 We use recent measurements of the expansion history of the universe to place constraints on the parameter space of cubic Galileon models, in particular we concentrate on those models which contain the simplest Galileon term plus a linear potential. This gives strong constraints on the Lagrangian of these models. Most dynamical terms in the Galileon Lagrangian are constraint to be small and the acceleration is effectively provided by a constant term in the scalar potential, thus reducing, effectively, to a LCDM model for current acceleration. The effective equation of state is indistinguishable from that of a cosmological constant $w=-1$ and the data constraint it to have no temporal variations of more than at the few \% level. The energy density of the Galileon can contribute only to about $10$\% of the acceleration energy density, being the other $90$\% a cosmological constant term. This demonstrates how useful direct measurements of the expansion history of the universe are at constraining the dynamical nature of dark energy.
\end{abstract}

\end{frontmatter}

\section{Introduction}

 The past decade in observational cosmology has been marked by the confirmation from different probes of the observed late-time accelerated expansion of the universe  \cite{Perlmutter:1998np,Riess:1998cb,Spergel:2003cb,Komatsu:2010fb,Hinshaw:2012aka,Eisenstein:2005su,Percival:2009xn,Ade:2013zuv}. The current challenge in theoretical physics and cosmology is to explain the nature of this acceleration. While the explanation as a pure cosmological constant is consistent with all data sets, other models that modify the Einstein-Hilbert action remain attractive as means of explaining acceleration. Among them, the Galileon models offer a robust framework in order to explain the dynamics of dark energy. They were originally introduced for a flat space-time \cite{Nicolis:2008in} in order to construct the most general single-field modified gravity theory which respects the Galilean-shift symmetry ($\pi\rightarrow\pi+b_\mu x^\mu +c$, with $b_\mu$ and $c$ constants) and avoids Ostrogradski instabilities (no more than second derivatives in the equations of motion). The generalization to a curved space-time, the covariant Galileon \cite{Deffayet:2009wt}, breaks softly the Galilean-shift symmetry, but avoids the Ostrogradski instabilities. 
 
In this model, the form of the action becomes
\begin{align} \label{EQ:ActionUncoupled}
S=\int {\rm d}^4 x \sqrt{-g}\,\left[ \frac{\Mpl^2}{2}R+\frac{1}{2} \sum_{i=1}^5 c_i \mathcal{L}_i \right]+\int {\rm d}^4 x\, \mathcal{L}_{M}\,,
\end{align}
where $c_{1-5}$ are dimensionless constants. $\mathcal{L}_{M}$ is the Lagrangian of a pressureless perfect fluid with density $\rho$ and four-velocity $u^\mu$, i.e. the dark matter. The five Lagrangian densities for the scalar field are
\begin{align}
\mathcal{L}_1=&M^3 \pi \\
\mathcal{L}_2=&(\nabla \pi)^2 \\
\mathcal{L}_3=&(\square \pi) (\nabla \pi)^2/M^3 \\
\mathcal{L}_4=&(\nabla \pi)^2 \left[2 (\square \pi)^2-2 \pi_{;\mu \nu} \pi^{;\mu \nu}-R(\nabla \pi)^2/2 \right]/M^6 \label{EQ:L4}\\
\mathcal{L}_5=&(\nabla \pi)^2 [ (\square \pi)^3-3(\square \pi)\,\pi_{; \mu \nu} \pi^{;\mu \nu}+2{\pi_{;\mu}}^{\nu} {\pi_{;\nu}}^{\rho}
{\pi_{;\rho}}^{\mu}+\nonumber\\
&-6 \pi_{;\mu} \pi^{;\mu \nu}\pi^{;\rho}G_{\nu \rho}] /M^9\label{EQ:L5}\,,
\end{align}
where $M$ is a constant with dimensions of mass and $\pi$ is the Galileon field. $\mathcal{L}_1$ is the most general potential term that respects the Galilean-shift symmetry in a flat space-time. $\mathcal{L}_2$ is the well known standard kinetic term. $\mathcal{L}_{3-5}$ are the so-called non-standard kinetic terms because they mix first and second derivatives of the scalar field. An important property of the Galileon models is the Vainshtein mechanism \cite{Vainshtein:1972sx}, which is due to the non-standard kinetic terms. This mechanism decouples the scalar field from gravity at small scales ($r\ll r_V$, where $r_V$ is a characteristic scale called Vainshtein radius), in order to satisfy solar-system constraints hiding the presence of a fifth force.

Even though the comparison of the Galileon with observations has already produced interesting results \cite{Nesseris:2010pc,Appleby:2012ba,Okada:2012mn,Barreira:2013jma,DeFelice:2011aa,Hirano:2010yf,Hirano:2011wj,Neveu:2013mfa}, in this paper we want to use the expansion history of the universe to constrain a subclass of these models: the cubic Galileon ($c_4=c_5=0$). In particular we take into account $c_1\neq 0$ which acts as a cosmological constant in the case $\pi'\rightarrow 0$. It is important to note that this condition can be reached only dynamically (see \cite{Bartolo:2013ws} for a discussion on the role of this term). Thus, with this setup we have a simple model that can eventually reduce to the $\Lcdm$ model. This will be particularly important in the parameter space analysis we are doing in the next sections, and it will affect our conclusions.


Throughout the paper we adopt units $c = \hbar= G = 1$; our signature is $(-, +, +, +)$. Greek indices run over $\{0, 1, 2, 3\}$, denoting space-time coordinates, whereas Latin indices run over $\{1, 2, 3\}$, labelling spatial coordinates.

\section{Galileon cosmology}

In a flat Friedmann-Lema\^{i}tre-Robertson-Walker (FLRW) universe,
\begin{align}
ds^{2}={a(\tau)}^2\left[-d\tau^{2} +\delta_{ij}dx^i dx^j\right]\,,
\end{align}
the Friedmann equations and the Galileon field equation read, respectively
\begin{align}
&\frac{3\Mpl^2 \mathcal{H}^2}{a^2}=\rho_m+\rho_\pi\,,\label{EQ:Friedmann1}\\
&\frac{\Mpl^2\mathcal{H}^2}{a^2}\left(1+ \frac{2a\mathcal{H}'}{\mathcal{H}}\right)=-p_\pi\,,\label{EQ:Friedmann2}\\
&\frac{c_1 M^3}{2}+ c_2\mathcal{H}^2\left[\pi''+\frac{\mathcal{H}' \pi'}{\mathcal{H}} +\frac{3 \pi'}{a}\right]-\frac{6c_3\mathcal{H}^4\pi'}{M^3 a} \left[\pi''+\frac{3\mathcal{H}' \pi'}{2\mathcal{H}} +\frac{\pi'}{a}\right]=0\,,\label{EQ:BackgroundGalileon}
\end{align}
where $\mathcal{H}\equiv a'(\tau)/a(\tau)$ is the Hubble parameter, primes represent derivatives with respect to the scale factor $a$ and
\begin{align}
\rho_{\rm \pi} \equiv& \frac{c_1 M^3}{2}\pi+\frac{c_2}{2}\mathcal{H}^2{\pi'}^2-\frac{3c_3}{M^3 a}\mathcal{H}^4 {\pi'}^3\,,\label{EQ:density}\\
p_\pi \equiv& -\frac{c_1 M^3}{2}\pi+ \frac{c_2}{2}\mathcal{H}^2{\pi'}^2+\frac{c_3}{M^3}\mathcal{H}^4{\pi'}^2\left[\pi'' +\frac{\mathcal{H}'\pi'}{\mathcal{H}}\right]\,,\label{EQ:pressure}
\end{align}
are the scalar field density and pressure, respectively. Since the mass scale $M$ can be easily absorbed into the coefficients $c_i$, without loss of generality, we have defined $M^3\equiv\Mpl \H0^2$, where $\H0$ is the value of the Hubble parameter $\mathcal{H}(\tau)$ today.

In principle, the background evolution, Eqs.\ (\ref{EQ:Friedmann2}) and (\ref{EQ:BackgroundGalileon}), of this model is given once six parameters $\{c_1,c_2,c_3,\mathcal{H}(a_i),\pi(a_i),\pi'(a_i)\}$ are fixed. In order to work with dimensionless quantities and to fix the initial conditions it is possible to renormalize the Hubble and the Galileon fields
\begin{align}
 \mathcal{H}(a)&\rightarrow h(a)\equiv \frac{\mathcal{H}(a)}{a \H0}\\
  \pi'(a)&\rightarrow x(a)\equiv \frac{a_i \pi'(a)}{a \pi'(a_i)} a^2 h(a)^2\,.
\end{align}

It is important to note that the background equations have a degeneracy in the parameter space. This means that different set of parameters can give the same cosmology. Thus, it is convenient to eliminate one degree of freedom (d.o.f.) through a redefinition of the parameters \cite{Neveu:2013mfa}
\begin{align}
 c_i &\rightarrow \,\, d_i\equiv {\left(\frac{\pi'(a_i)}{a_i\Mpl}\right)}^i c_i\,.
\end{align}
It is possible to use the first Friedmann equation, Eq.\ (\ref{EQ:Friedmann1}), to eliminate the potential term in Eq.\ (\ref{EQ:Friedmann2}) \cite{Easson:2011zy,Easson:2013bda}. Together with Eq.\ (\ref{EQ:BackgroundGalileon}), and using this reparametrization, we have now our set of two first-order differential equations
\begin{align}
&2 h h'= -3 \Omega_{m0} a^{-3}-\frac{d_2 x^2}{h^2} - \frac{d_3 x^2}{h^2} \left(x'-3 x- \frac{x h'}{h}\right)\label{EQ:solve1}\\
&\frac{d_1}{2} +d_2 \left(x' - \frac{x h'}{h} +3 x \right)-6 d_3
   x\left(x'- \frac{x h'}{2 h} + \frac{3}{2} x\right) = 0\,,\label{EQ:solve2}
\end{align}
where $\Omega_{m0}\equiv \rho_{m0}/(3\Mpl^2 {\mathcal{H}_0}^2)$ is the value of the matter density today. The parameter region we have explored satisfies \cite{Appleby:2011aa}
\begin{align}
&d_2-6 d_3 x+ \frac{3 {d_3}^2 x^4}{2 h^4}>0\\
&d_2 - 2 d_3 a \left(x'+\frac{2 x}{a}-\frac{x h'}{h}\right) - \frac{{d_3}^2 x^4}{2 h^4}\geq 0
\end{align}
in order to avoid ghost and laplace instabilities.

The initial conditions for the Galileon field is $x(a_i)={a_i}^2 h^2(a_i)$, while we don't need to fix $\pi(a_i)$. Even if the initial conditions are completed by fixing $h(a=1)=1$, we found it is convenient to add another constraint in order to avoid instabilities in the numerical integration of the differential equations, precisely we imposed that at early times the energy density of DM was dominand w.r.t. the energy density of the Galileon
\begin{align}
h^2\simeq \frac{\Omega_{m0}}{a^3}\,.
\end{align}
Therefore, the set of parameters we need to study the background dynamics, Eqs.\ (\ref{EQ:solve1}) and (\ref{EQ:solve2}), will be $\{\Omega_{m0},d_1,d_2,d_3\}$.

\section{Numerical Results}

\begin{figure}[!ht] 
   \centering
\includegraphics[width=0.8\textwidth]{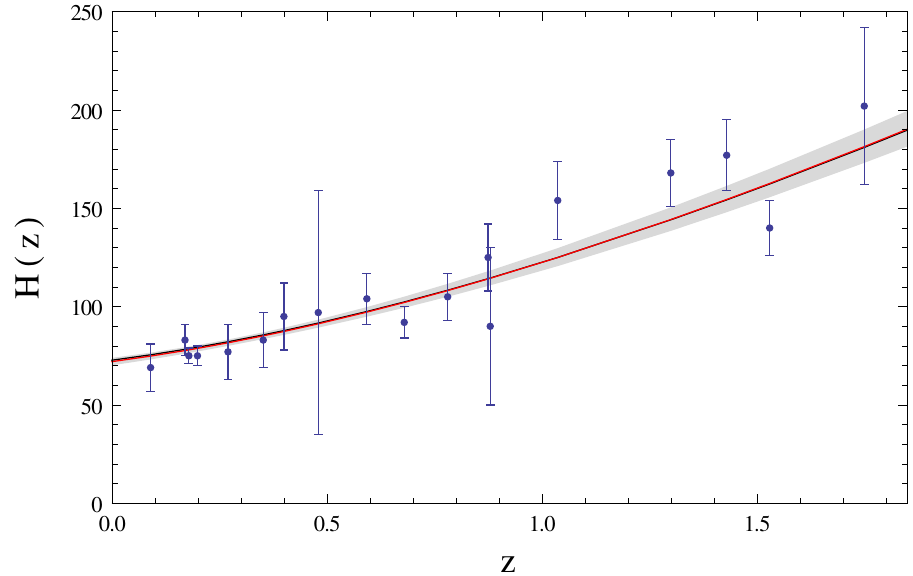}
\caption{Hubble parameter $H$ as a function of the redshift $z$. The blue points are the data we use in our analysis. The black line shows the evolution of the best fit $\Lcdm$ model with $\H0=72.8$ and $\Omega_{m0}=0.26$. The red line shows the best fit we have found for the Galileon evolution, i.e.\ $\{\mathcal{H}_0\simeq 72,\Omega_{m0}\simeq 0.27, d_1\simeq -47, d_2\simeq 13\times 10^3, d_3\simeq -14\}$. The grey area shows the 1-sigma region for the joint distribution of the parameters in the Galileon Lagrangian.}\label{FIG:Hz}
\end{figure}

Age measurements of massive, red galaxies can be used to estimate the upper edge of the age distribution at each redshift, the so-called red envelope ages. These measurements of the oldest galaxy ages vs.\ redshift can be used as a redshift-dependent lower bound on the age of the universe. In total we use $32$ such age estimates in the redshift range $z = 0.1 - 1.85$, with independent error bars at the 10\% level (see Fig. \ref{FIG:Hz}). We refer to \cite{Jimenez:2001gg,Simon:2004tf,Stern:2009ep,Moresco:2010wh,Moresco:2012jh} and references therein for details on the data sets used and on the age estimation from galaxy spectra. In particular, the difference between our work and the interesting results obtained in \cite{Kimura:2010di,Ali:2010gr} is that we are directly constraining $H(z)$ rather than its integral as the SN Ia distances.

Using these data we have minimized the $\chi^2$ distribution
\begin{align}
 \chi^2=\sum_{i=1}^{\mathcal{N}}\frac{{\left[\H0 h(z_i) -H_{obs}(z_i)\right]}^2}{{\sigma_i}^2}\,,
\end{align}
with a Nelder-Mead algorithm. From now we shall introduce another variable (i.e. $\H0$, which of course does not influence our background equations of motion), leading to a five-dimensional parameter space. Trying with different initial points, we have found that the best fit has $\chi^2\simeq 13.9$, and its coordinates are $\{\mathcal{H}_0\simeq 72,\Omega_{m0}\simeq 0.27, d_1\simeq -47, d_2\simeq 13\times 10^3, d_3\simeq -14\}$.

\begin{figure}[!ht] 
\includegraphics[width=0.5\textwidth]{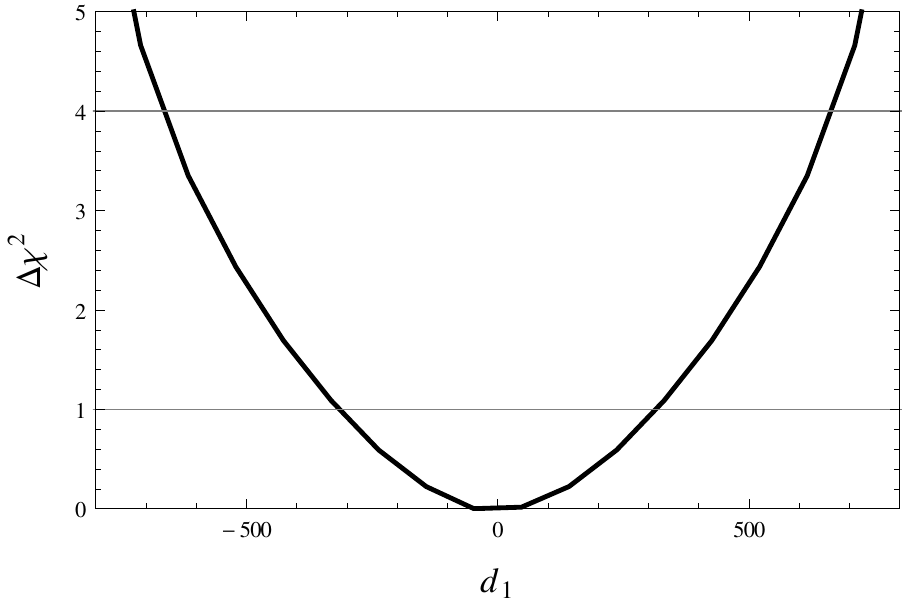}\includegraphics[width=0.5\textwidth]{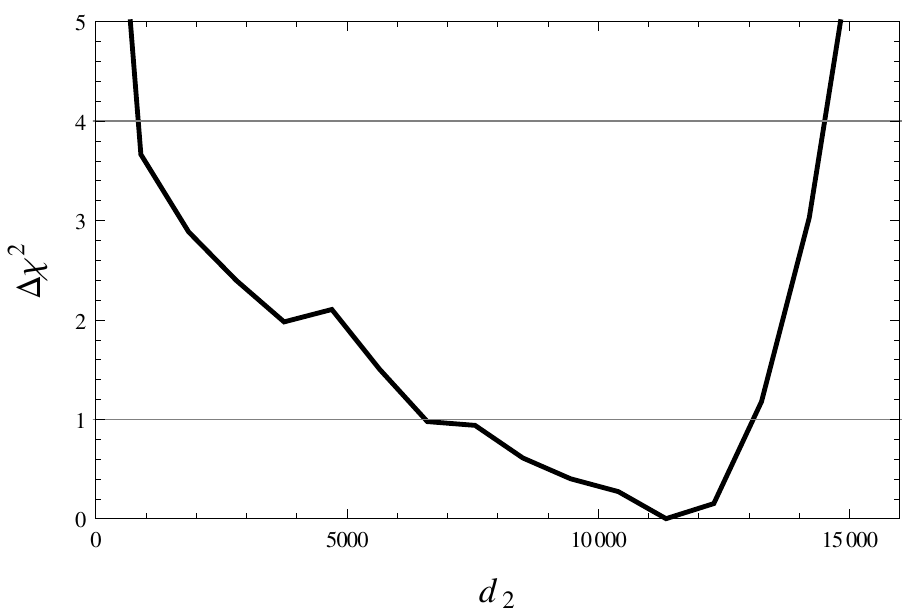}\\
\includegraphics[width=0.5\textwidth]{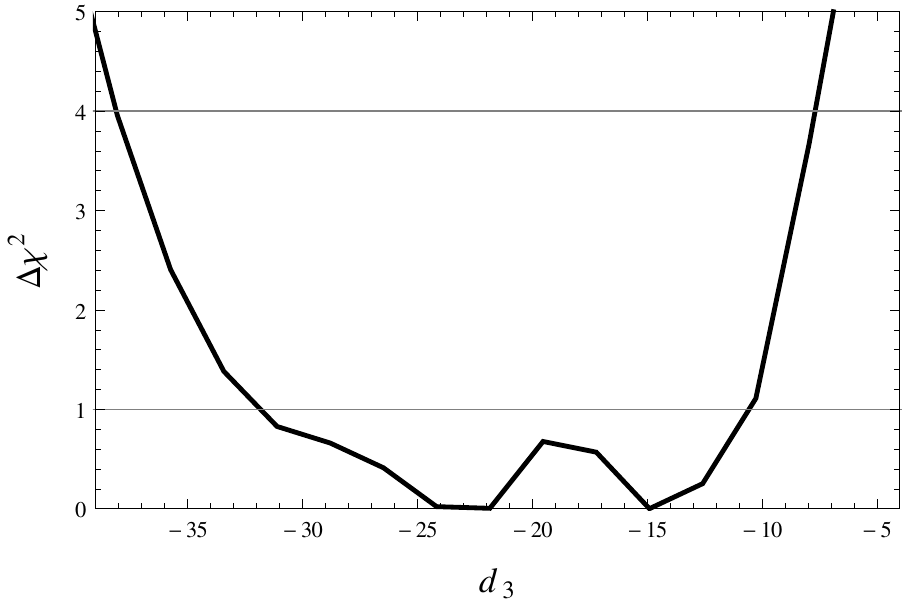}
\caption{$\Delta\chi^2$ for the marginalised distribution of each parameter. Black thick lines refer to the distribution obtained using a Gaussian prior for $H_0$ ($H_0=73.8\pm 2.4\, {\rm km\, s^{-1}\, {Mpc}^{-1}}$), while the horizontal lines represent the $1\sigma$ ($\Delta\chi^2=1$) and the $2\sigma$ ($\Delta\chi^2=4$) bounds for the 1D marginalised distributions.}\label{FIG:best1}
\end{figure}

\begin{figure}[!ht] 
\includegraphics[width=0.5\textwidth]{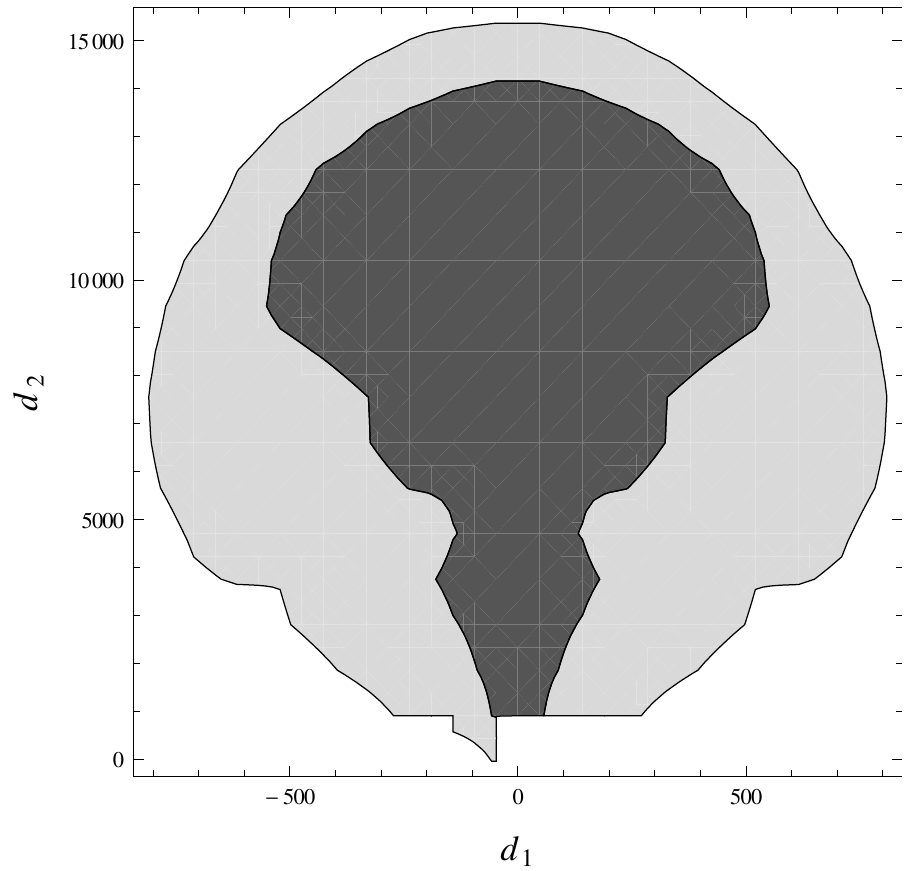}\includegraphics[width=0.5\textwidth]{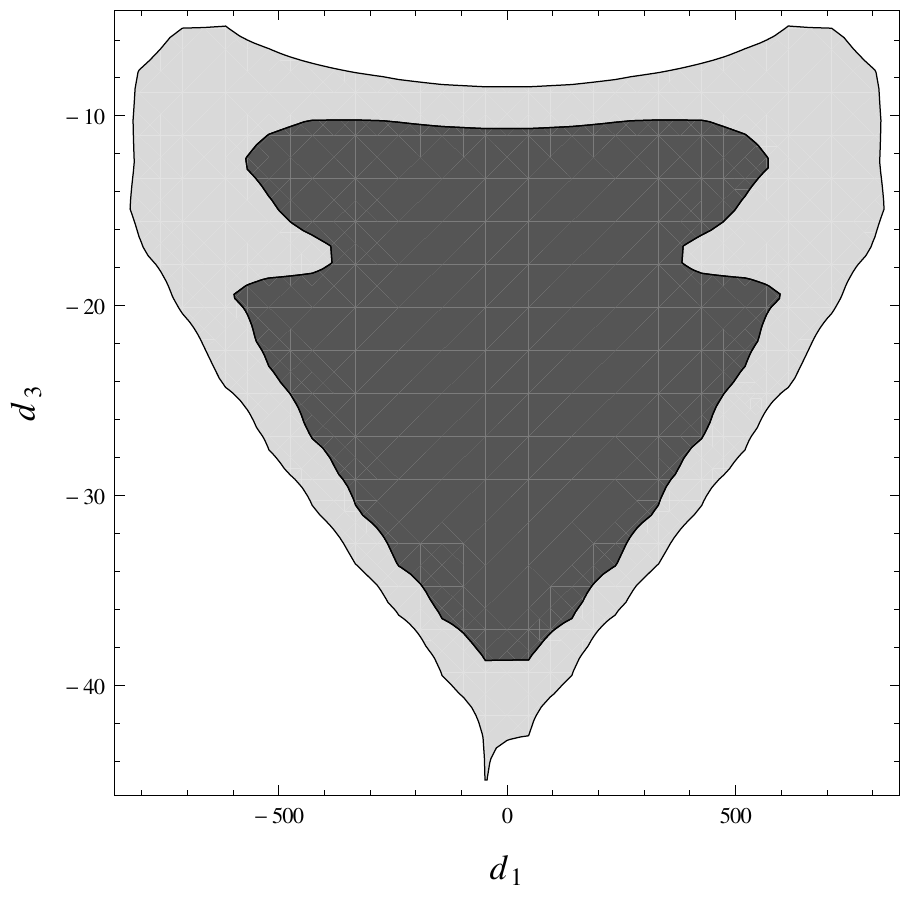}\\
\includegraphics[width=0.5\textwidth]{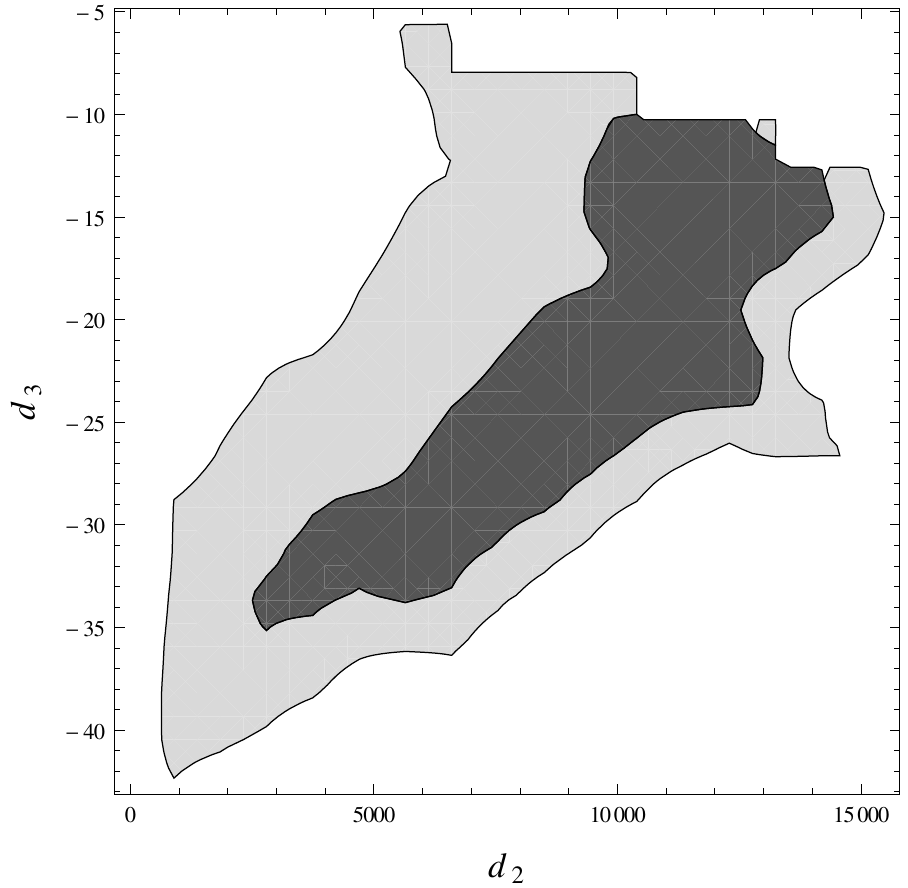}
\caption{$1\sigma$ (dark gray) and $2\sigma$ (light gray) regions for the joint distribution in 2D. In all the panels we have used a Gaussian prior on $H_0$ ($H_0=73.8\pm 2.4\, {\rm km\, s^{-1}\, {Mpc}^{-1}}$).}\label{FIG:best2}
\end{figure}

We then explored the parameter region around this minimum with grids of 50 points in each dimension, in order to plot the $1\sigma$ ($68.3\%$ CL) and the $2\sigma$ ($95.4\%$ CL) regions. In particular, in Fig.\ \ref{FIG:best1} we plot the 1D marginalised distributions for each parameter. In Fig.\ \ref{FIG:best2} we plot the $1\sigma$ ($\Delta\chi^2=2.3$) and the $2\sigma$ ($\Delta\chi^2=6.18$) regions for the 2D joint distributions. In both figures we show the results obtained using a Gaussian prior on $\H0$ ($\H0=73.8\pm 2.4\, {\rm km\, s^{-1}\, {Mpc}^{-1}}$) \cite{Riess:2011yx} and a flat prior on $\Omega_{m0}\in [0.26;0.30]$.

In Figs.\ \ref{FIG:best1} and \ref{FIG:best2}, it can be seen that the parameter $d_2$ can be significantly larger than the other parameters. This result is expected, indeed during the matter-dominated epoch we have the following approximated relations
\begin{align}
 \Omega_m(a\ll 1)&\simeq 1\\
  \Omega_\pi(a\ll 1)&\simeq \frac{d_1}{6\Omega_{m0}} a^3 + \frac{d_2}{6} a^4 - d_3 \Omega_{m0} a^3\,.
\end{align}
Here, the $d_1$ and $d_3$ terms scale as $\propto a^3$, while $d_2$ as $\propto a^4$. This means that the Galileon initial energy density is determined mostly by the $d_{1,3}$ terms, while the $d_2$ term can be increased by a factor $a^{-1}$ ($10^3$ at our initial time) before affecting the dynamics at early times. If we want to generalize this statement to the epochs in which the Galileon contribution becomes non-negligible, we have to consider the behavior first noted in \cite{Appleby:2011aa}, when the authors describe the hierarchical dynamics of $\rho_\pi$ (the difference is that they are not considering $d_1$). Taking into account all the $d_{2,5}$ terms, it is shown that if $d_i$ is dominant at a certain epoch, the following terms (i.e.\ $d_{i+1}$, $d_{i+2}$, \ldots) will remain subdominant at all subsequent times.

\begin{figure}[!ht] 
\includegraphics[width=0.5\textwidth]{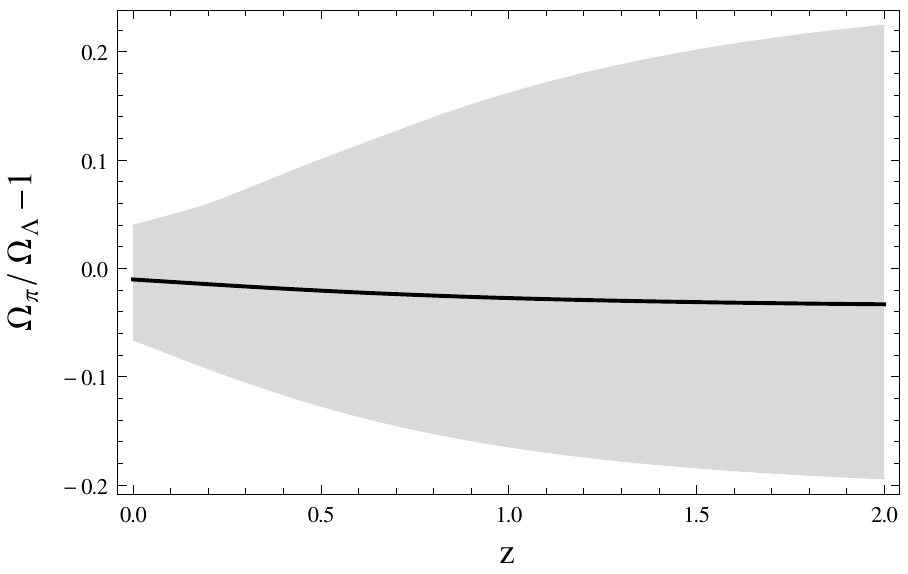}\includegraphics[width=0.5\textwidth]{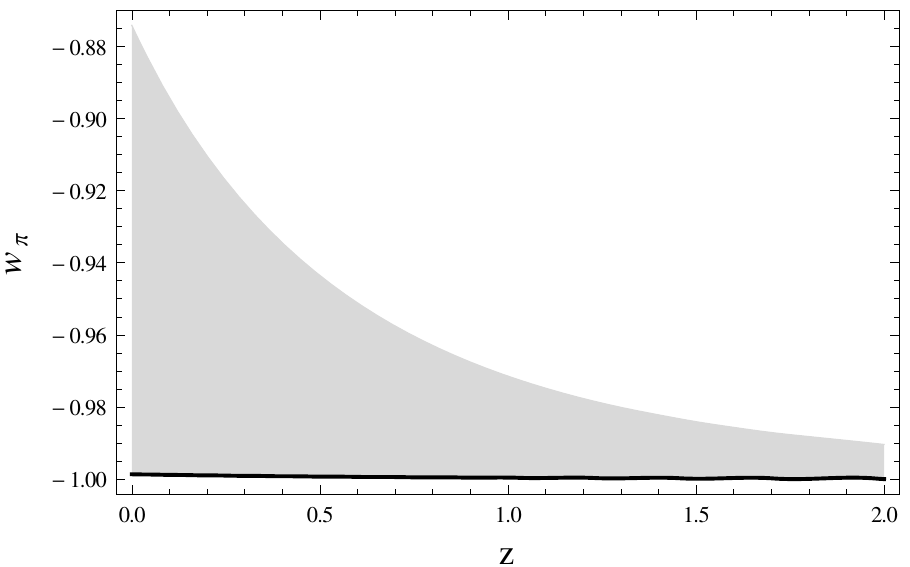}
\caption{Left panel: the evolution of the Galileon energy density ($\Omega_\pi$) w.r.t. the evolution of the $\Lcdm$ model energy density ($\Omega_\Lambda$). Right panel: the evolution of $w_\pi$ as a function of the redshift $z$. Solid lines are the best fit model while gray areas represent the $1-\sigma$ region for the joint distribution.}\label{FIG:omegawbest}
\end{figure}

Having obtained constraints on the coefficients of the Galileon Lagrangian, we turn our attention to the meaning of this results in more observational terms. In Fig.~\ref{FIG:omegawbest} we show the values of $\Omega_{\pi}$ and $w_{\pi}$ for the best fitting parameters to the $H(z)$ data (solid line) and their $1-\sigma$ uncertainty regions, obtained from the joined distribution of $d_{1-3}$. The best fitting value of $w_{\pi}$ is indistinguishable from that of a cosmological constant at the 0.1\% level. The $1-sigma$ range allows for variations only of few \% from the value of a cosmological constant. The relative contribution of $\Omega_{\pi}$ to $\Omega_{\Lambda}$ is $\sim 1$ at the $10$ \% level, while for the best fitting model is indistinguishable from a cosmological constant term. Even within the $1-\sigma$ regions nearly $90$\% of the accelerating energy density has to be a cosmological constant. These two parameters indicate that the dynamics of the Galileon is nearly inexistent and that it behaves mostly as a cosmological constant. A similar result was found in \cite{Leon:2012mt}, where the authors perform a dynamical analysis in the context of the Generalised Galileon (or Horndeski) theory. In addition, in \cite{Ali:2012cv}, the authors claim that the background evolution of the cubic Galileon with a potential is not significantly affected by $\mathcal{L}_3$. This result agrees with our observation that for viable models the background dynamics is driven by the potential term.

\section{Conclusions}

In this short note we have shown constraints obtained by comparing the measured expansion history of the universe and the prediction from the cubic Galileon model with a linear potential. We have found tight constraints in most of the Lagrangian terms of the model. Even with the addition of 3 extra free parameters, the best fit we have found has $\chi^2\simeq13.9$ vs.\ ${\chi_\Lambda}^2\simeq16.0$, which is not a significant improvement. In fact, using a simple bayesian evidence computation the Galileon model is excluded at the "Decisive" level (odds $> 100:1$ against the Galileon model). This conclusion is also supported by exploring the cosmological observables $\Omega_{\pi}$ and $w_{\pi}$, which indicate a behaviour similar to a cosmological constant for the model. The expansion history measurements have proven extremely useful at constraining the dynamical evolution of dark energy (see also Ref.~\cite{Jimenez:2011nn,Jimenez:2012jg} where we constrained the dynamics of an effective general dark energy Lagrangian to be less than at the $7$\% level). Future measurements at the \% level of the expansion history of the universe from the Euclid satellite, will provide an even more stringent test on the dynamics of recent cosmic acceleration.

\section*{Acknowledgements}
We thank Valerio Marra for useful discussions. We thank Imperial College and in particular the ICIC centre for generous funding under the ``CosmoClassic" initiative that served to start this collaboration. EB would like to acknowledge the ``Fondazione Ing.\ Aldo Gini'' for support during part of the development of this project.

\end{document}